# Heidegger's Quantum Phenomenology

François-Igor Pris[1]

**Abstract**

The article suggests that quantum mechanics is a science of a new type, which refutes the classical metaphysical concept of reality. The notion of a quantum concept is introduced. The possibility of a Wittgensteinian "dissolution" of the measurement problem with the help of the notion of a language game and the possibility of a metaphysical solution of this problem with the help of the Heideggerian notion of *Dasein* are considered.

## The Measurement Problem and Quantum Concepts

Some philosophers think that the measurement problem is the principal philosophical problem of quantum mechanics (see, for example, Wallace 2008). This problem has been widely discussed in the literature since the creation of quantum mechanics. Many different solutions to the problem have been proposed. Until now though, no consensus has been reached.

In this paper I provide some reasons supporting the claim that Heidegger's philosophy (Heidegger 1967, 1996) contains resources that allow one to better understand the measurement problem and to come closer to its solution, if not to solve it. Moreover, in a sense Heidegger's phenomenology is the appropriate philosophy for the understanding of non-classical physics in general.

1 Francois has a doctorate in quantum mechanics and is a researcher at the Universität Dortmund, Germany.

The quantum theory is considered as a non-classical theory, that is, as a theory of a principally new type, while the Einstein theory of relativity, despite its revolutionary character, is considered as a classical theory. What is the difference between classical and non-classical physics?

Let us adopt the realistic point of view and suppose that science investigates reality. Then the border between classical science and non-classical science can be drawn in accordance with the concept of reality which is being used.

Any concept of reality supposes its "objectivity", that is, the independence of what is real from the subject. But this independence can be understood differently.

The concept of reality, which is implicit or explicit in classical physics, is what philosophers call metaphysical realism. According to this concept, reality of things, facts and phenomena does not depend on the subject in a certain absolute sense; it can be completely detached from the subject and opposed to her as an "exterior world". The subject-scientist learns about this reality by means of a theoretical (mathematical) representation; she is situated, so to say, face to face with the reality and theoretically represents or "reflects" it. The concepts being used are those of the classical type, in the sense that the result of their application is always predetermined. For instance, the use of the classical concepts of coordinates and momentum allows one to determine the coordinates and momentum of a particle, which have definite values even when the concepts of coordinates and momentum have not been used.

Gary Ebbs formulates the idea of metaphysical realism as follows (Ebbs 1997, p. 203; cited in Wilson 2008, p. 79):

"The idea behind metaphysical realism is that we can conceive of the entities and substances and species of the 'external' world independently of any of the empirical beliefs and theories we hold or might hold in the future."

According to French philosopher-phenomenologist Jocelyn Benoist, the classical (metaphysical) realism is characterized by two traits: (1) there is, in an important sense, (objective) reality which is independent from the subject; (2) this objective reality is the reality of objects situated in front of the subject. Benoist keeps (1), but rejects (2). For him, the genuine reality is the reality of interaction between the subject and the world. (Benoist 2005)

Heisenberg distinguishes between dogmatic realism, which, according to him, is the point of view of classical physics, metaphysical realism and practical realism. The latter is the natural realism of science. The dogmatic realism claims that all meaningful affirmations about the material world can be made objective. The metaphysical realism is the dogmatic realism together with the claim that things really do exist. The notion of metaphysical realism according to Heisenberg is thus the traditional notion of metaphysical realism (conditions (1) and (2) above).

Heisenberg writes that Einstein criticized quantum mechanics from the point of view of dogmatic realism. (Heisenberg 1989, pp. 43 – 45.) In reality, however, Einstein's position is much more nuanced. For instance, he writes: "Physics is an attempt conceptually to grasp reality as it is thought independently of its being observed" (Einstein 1949, p. 81; cited in Stapp 1998). This corresponds only to the condition (1) above.

The metaphysical realism leads to the difficulties in interpretation of quantum mechanics, which can be dissolved if one replaced it by a non-traditional form of realism.

Although within quantum mechanics the notions of subject and object are meaningful, the ultimate quantum mechanical reality is that of the "process of measurement" in which the subject interacts with the object and in this interaction it is as though both become "dissolved", inseparable from each other.

Before the measurement, it makes sense to talk about a quantum system situated in a state, for example, in a state of superposition of *eigenstates* of a physical quantity (of a *Hermitian* operator) and also about the subject, or observer, which is independent of the quantum system and does not interact with it. After the measurement, there is also an independent subject-observer and a quantum system which is independent of the subject, and which is situated in one of the *eigenstates* of the measured physical quantity. After the act of measurement, the physical quantity has a definite value. The result of its measurement is reproducible. But in the process of measurement, when the probabilistic reduction of the wave function takes place, it makes no sense to talk about the subject as such and the object as such. The process of measurement (reduction of wave function) cannot be called a physical process in the usual sense. It is unobservable and cannot be mathematized, even in principle. The Born rules determining the probabilities of transition from a superposition state into one of the *eigenstates* of a physical quantity, i.e. the probabilities of obtaining a definite result of measurement only establishes a correspondence between the initial situation and the final situation and does not say anything about the "process" of transition itself.

The principal hypothesis of this paper, in favour of which some arguments are given below, is that it has a phenomenological nature in the sense of the Heideggerian *Dasein*.

Quantum theory can be used as an instrument for preparing a given experimental situation (it seems that this demands more reflection on the foundations of the theory and the nature of reality than in the case of classical physics), but instrumentalism without phenomenology is not an appropriate philosophy of quantum physics. (See also the following historical accounts (Carson 2010a, 2010b, 2010c).)

The result of the application of quantum concepts, for example, of the concepts of coordinates and momentum of a particle, is not predetermined: it is false or even meaningless to speak about the simultaneous existence of definite values of position and momentum of a particle or about the existence of a well-defined trajectory of a quantum particle. Quantum concepts are rules (operators) for obtaining definite values. A definite value appears only as the result of an application of a quantum concept, that is, as the result of a measurement. For example, if the momentum of a free particle has a definite value, its position is absolutely undetermined. In this case the result of an application of the concept of a coordinate is absolutely undetermined (though quantum probabilities are determined): the coordinate with equal probability can take any value; a definite value appears only as the result of a process of measurement of the coordinate.

Classical (commutative) physical quantities represent some real numbers. This cannot be said about the corresponding quantum operators. For a given quantum state, they represent matrices (sets) of possible values of physical quantities together with the corresponding probabilities. The actualization of a given definite value happens during the process of measurement.

One can say that quantum concepts of physical quantities represent some quantum physical quantities (properties), which should be understood as dispositional ones. (About dispositional interpretation of quantum mechanics see, for example, works by Suarez (2004, 2007).) In (Suarez 2004, p. 233, footnote 12), Suarez writes that the representation of a quantum property has no analogue in quantum mechanics:

"There is no analogue of this type of presentation in classical mechanics. (…) A quantum state is not to be interpreted à la classical mechanics as assignments of actually possessed properties and their values, but rather as a mere assignment of probabilities."

Although before the measurement of the position of the electron, the electron does not have any definite position, one cannot say that it does not have any position at all: the position of the electron (given its wave function) is a real dispositional property, which is described by the operator of its position.

That is why for Heisenberg "(...) the atoms or the elementary particles (...) form a world of potentialities or possibilities rather than one of things or facts" (Heisenberg 1958, p. 160; cited in Suarez 2007, c. 423, footnote 8).

In spite of its revolutionary character, the theory of relativity (special and general) is considered as a classical theory precisely because the concepts used by the theory (and the theory as a whole) function in the classical regime. The observer measures the quantities whose concrete values exist before the measurement (though they depend on a reference frame), i.e., independently of whether or not the observer produces a measurement. The Einstein principle of relativity saying that there is no privileged physical framework (different observers observe different values of a physical quantity) does not take into consideration the relativity of the border between the observer and what is observed. In quantum mechanics this border (in classical sense) is fixed only *post factum*, as the result of a process of measurement.

A classical concept and the corresponding quantum concept can be understood as two aspects of one and the same more general concept.

According to Mark Wilson, even "simple" concepts have a rich internal fine structure consisting of sub-concepts related to each other in different ways. (Wilson 2008)

Some concepts function as an atlas consisting of various partially overlapping leaves (maps). Here is one of his examples, cited by Robert Brandom (Brandom "Platforms…", p. 6):

"Mass, impressed gravitational force, and work required to move something relative to a local frame are (some of the) leaves of the atlas-structured empirical concept weight."

In the case of the "atlas" structure there is a Wittgensteinian family resemblance between the different "leaves" of an atlas.

A more complex conceptual structure is the "patchwork" one, proposed by Wilson. The various leaves in a patchwork are connected to each other at their edges. Such is, for example, the concept of hardness. Brandom describes this example as follows (Brandom "Platforms…", p. 6):

"Hardness generically is something like resistance to penetration. To test such resistance, we might press a weight on a sample, squeeze it, strike it, scratch it, cut, or rub it. The results of these various tests will not always be consilient."

The generalized concepts of physical quantities representing classical as well as quantum properties can be understood as "atlas-patchwork" ones. The domains of application of a classical concept and the corresponding quantum concept are overlapped when the physical quantity has a definite value. (Wilson formulates 44 principal theses characterizing classical concepts (see Wilson 2008, pp. 139-146). His understanding of the relation between "classical" and "quantum" concepts is somewhat different from our own (Wilson 2008, p. 197).)

**The Wave Function And Quantum Reality**

Let us illustrate the question about the quantum reality, quantum objectivity and quantum concepts by using the example of the wave function.

It is known that the wave function suffers from the lack of objective reality in the following sense (see, for example, (Haroche 2006)). A wave function can be determined in a statistical context, when there is the possibility of performing measurements on its identical copies. If one deals with only one exemplar of a quantum system situated in a pure state ψ, which is an *eigenstate* (with the *eigenvalue* 1) of a projector, the function ψ is known only to the experimenter who prepared it, but cannot be known to an exterior observer. Indeed, if an observer performs a measurement, the system modifies its state irreversibly and randomly. A part of the information about the initial state of the system is lost. The experimenter obtains only partial information about it. (As a consequence, it is impossible to copy exactly an unknown quantum state (the no-cloning theorem).)

Hence one should either accept the existence of unknowable reality (unknowable from the point of view of the observer who is not implicated in the process of preparing the wave function), or recognize that the ultimate reality supposes the presence of the interacting subject-observer: independently of the observer, the quantum function cannot be considered as objectively real; the quantum reality is the reality of the act of preparing the state of the system.

In the last case, the concept "wave function" cannot be viewed as a concept in the classical sense, but as a "quantum concept" whose functioning supposes the presence of the observer.

In substance, in quantum mechanics, concepts are functioning not as a means for representing a metaphysical reality of objects, existing independently from the observer, but as rules for the interaction of the observer with the reality, for the formation of the classical "metaphysical reality", which is secondary, by the observer. Hence the understanding of quantum mechanics should not consist of understanding quantum concepts in a classical way, which is not possible, but in understanding that such understanding is not possible, that quantum concepts differ from the classical ones in the way they function.

What has been said agrees with the following position which Wilson (2006, 2008, p. 77) attributes to Kant:

« … The general claim is that our naive conception of « objective » concepts as *corresponding* to real world attributes is incoherent; that every viable concept must inherently involve the constructive agencies of our own minds in some irrevocable way"

## The Measurement Problem, the "Hard Problem" in the Philosophy of Mind and the Wittgensteinian Problem of Rule-Following

The act of measurement – action of the subject in the process of measurement – is being performed within an established scientific and ordinary practice, in accordance with implicit and explicit rules. That is why it is possible to solve/dissolve the measurement problem pragmatically.

Michel Bitbol, for example, proposed a Wittgensteinian "dissolution" of the measurement problem within the second philosophy of Wittgenstein (without making an appeal to Heidegger's philosophy). The dissolution consists of using the instruments and mathematical symbols in a way so that the problem no longer appears. (Bitbol 2000a. See also Bitbol 2000b, 2002, 2008.)

This is not a naive avoidance of attempts to solve the problem. The Wittgensteinian "dissolution of the problem" is the initial point but also the final point of a long series of failed attempts to solve the problem formally or discursively.

Bitbol emphasizes that the quantum observation is not an observation of pre-existing entities. It is a practice organized according to procedural rationalities guided by theoretical rules. The ontology in the sense of Quine is a secondary retranslation of this practice. The act of measurement is an execution of procedures which, for a certain class of well-defined experiential preparations, gives reproducible values. These values can be treated as secondary, as reflecting properties of objects.

According to Bitbol, the function of "I" is to manifest an engagement in the double sense of this word: engagement to accomplish something and engagement in a situation. The formalism of quantum mechanics taken in isolation from the practice of its application is incomplete, and a formal completion of quantum mechanics is impossible. Nevertheless, says Bitbol, a larger system including the quantum formalism, the probability rules of its application and its effective confrontation with the achievements of every concrete experiential situation has been complete since the creation of quantum mechanics (a performative completeness). (Bitbol 2000a, p. 342.)

Though at first sight no two philosophers are more different than Heidegger and Wittgenstein, there is no doubt that both are pragmatists. Brandom, for example, understands Heidegger's philosophy as a normative pragmatism. (Brandom 2002.) The normative social practice is primary. Norms and rules are implicit in practice. Phenomena, objects and the subject herself are secondary: they can be (re)constructed pragmatically. (Brandom states his theory of normative pragmatism in Brandom 1994, Brandom 2000.)

One can agree with Rouse (2002) that Brandom's anti-naturalistic interpretation of Heidegger should be turned on its head. Heidegger is a naturalist. Natural phenomena are primary. They contain in themselves their own implicit norms.

It is pretty obvious that the second philosophy of Wittgenstein is also a specific (normative) naturalism. (Pris 2008.) The Wittgensteinian language games, governed by the implicit or explicit natural rules, are both natural and spontaneous (normative). Moreover, there are some reasons to think that Heidegger's metaphysics is implicit in Wittgenstein's philosophy and the Heideggerian notion of *Dasein* (literally Being-there) corresponds to the Wittgensteinian notions of language game and form of life.

For example, the following characterization of Dasein is also applicable to the language games (Heidegger 1996, ch. 2, § 12):

"Dasein is an entity which, in its very Being, comports itself understandingly towards that Being. Dasein exists. Furthermore, Dasein is an entity which in each case I myself am. Mineness belongs to any existent Dasein, and belongs to it as the condition which makes authenticity and inauthenticity possible."

According to Dreyfus, in contrast to Wittgenstein, Heidegger needs in a special technical philosophical language to theorize the background, i.e. practices. (Dreyfus 1991.) In other words, Heidegger's phenomenology makes explicit what is implicit in multiple Wittgensteinian examples. The philosophical literature indicates the existence of a close relationship between the notions of language game and form of life and Dasein. (As for the relations between Wittgenstein and Heidegger see, for example, (Mulhall 1990), (Weston 2010), (Dreyfus 1991), (Rentsch 2003 and the bibliography therein), (Egan 2013), (Benoist 2010), (Laurent 2015).) Heidegger himself writes that "Language is not identical with the sum total of all the words printed in a dictionary; instead… language is as Dasein is … it exists." (Heidegger 1996.)

That is why one can suppose that together with the Wittgensteinian notion of language game, the Heideggerian notion of Dasein can be used for pragmatico-phenomenological solution of the measurement problem in quantum mechanics. To be more precise, the notion of language game rather allows a therapeutic "dissolution" of the measurement problem. The notion of Dasein allows an explicit metaphysical solution of it (see below).

The theoretical rules used by the subject-observer for explanations and predictions of quantum-mechanical phenomena, and the real material quantum mechanical system – the things rules are being applied to – are two complementary aspects of the functioning of quantum mechanics. During the process of measurement, one or another possibility is being actualized. The "gap" between the phase of a theoretical prediction and the phase of asserting a result of measurement, that is, between the domain of possible and that of actual, is being closed pragmatically (see, for example, (Bachtold 2009)).

Hence the measurement problem can be understood as an instantiation of the Wittgensteinian problem of rule-following, or as the problem of a "gap" between the quantum mechanical rules, or concepts, and their application, where the role of the quantum observer (thus of consciousness) is significant. This problem can be "solved" or "dissolved" à la Wittgenstein: the process of measurement is a Wittgensteinian "process" – a language game – of the application of quantum rules.

The so-called explanatory gap problem in the philosophy of mind, which is also called the "hard problem" of consciousness, that is, the problem of a physicalistic or naturalistic explanation of a phenomenal consciousness, can also be understood as a Wittgensteinian problem of a "gap" between a rule (concept) and its application.  Hence the measurement problem in quantum mechanics is an instantiation of the explanatory gap problem, or the hard problem, in the philosophy of mind.

The existence of different kinds of dualistic solutions of the measurement problem: the affirmation that the reduction of the wave function is due to the consciousness of the observer; or, on the contrary, that it generates consciousness; the Wigner solution of the measurement problem, which makes appeal to the consciousness of the observer; the "many-minds" solution, and so on, indicates that it is necessary to take consciousness into account. At the same time, it is a consequence of a reification of consciousness which is not a "nothing", but is not a "something" either - a paraphrase of the Wittgensteinian claim about sensations. (According to Bitbol Wittgensteinian's idea is that "consciousness" should be taken as immediate experience rather than self-awareness. (Bitbol 2008)): «Sie ist kein *Etwas*, aber auch nicht ein *Nichts*!» (It is not a *something*, but not a *nothing* either!) (Wittgenstein 1953, 2001, §304). Although Wittgenstein speaks about sensations, for him this is also true for consciousness in general.

Consciousness must be included in quantum mechanics and physics in general, not as a non-physical substance or non-physical properties, but as a primary Given – an immediate experience, which in itself is physical. This is the beginning and the end of a theory. The immediate experience closes the explanatory gap between a theory and its application. The "hard problem" in the philosophy of mind, or the problem of closing the "gap" between phenomenal consciousness and its physicalistic description, appears as a result of reification of consciousness. It can be "solved" or "dissolved" only by means of a correct understanding of the naturalistic (physicalistic) nature of consciousness.

So, briefly, the logic of a pragmatic "dissolution", or normative-naturalistic solution, of the measurement problem is the following:

(1) The measurement problem is a particular case of a more general problem: the hard problem, or the problem of the explanatory gap, in the philosophy of mind.  (Michel Bitbol (2000) writes that the measurement problem and the hard problem have one and the same structure.) A human being, and thus consciousness, is essentially involved in the process of measurement: a human being forms an intention to accomplish a measurement, and accomplishes it, and finally observes the result of the measurement. (Heelan (2004), for example, defends the thesis that consciousness has, on phenomenological grounds, a similar structure to quantum mechanics.)

(2) The hard problem can be reduced to the problem of application of a rule (concept or theory) in the Wittgensteinian sense. (Pris 2008.) The explanatory gap is a "gap" between a neurological concept, describing a phenomenological experience and its application to the phenomenological experience; hence it is also a "gap" between a neurological concept and the corresponding phenomenological concept. (A Wittgensteinian "dissolution" of the hard problem was also proposed by Bitbol (2000a), but not in terms of the Wittgensteinian problem of rule-following.)

(3) The act of application of a rule is a Wittgensteinian language game. Within the language game the explanatory gap is closed.

(4) The notion of language game can be understood not only pragmatically, that is as a normative activity, interaction, or practice, but also naturalistically, that is, as a natural phenomenon. The language game is both natural and spontaneous (normative). Hence the Wittgensteinian naturalism is not metaphysical, but normative, which means that the phenomenon contains in itself its own implicit norms that are themselves natural. (The corresponding notion in Joseph Rouse – "phenomenon", or "intra-action". According to Rouse, we should stop conceiving of the normative and the material as separate – the two are constitutive of each other. (Rouse 2002.))

(5) Heidegger's notion of Dasein corresponds to Wittgenstein's notion of language game.

Thomas Rentsch (2003) thinks that Dasein corresponds rather to the Wittgensteinian notion of form of life. However, the Wittgensteinian notion of a form of life can be understood as a system of established language games, background, or as a sort of "language game of the second order" within which it only makes sense to consider more concrete language games of the first order.

(6) The so-called collapse of the wave function in the process of measurement is not a physical process. The process of measurement in quantum mechanics is a Wittgensteinian language game, or, in metaphysical Heideggerian language, Dasein (or a realization of one of the possibilities of Dasein).

Notice in this connection that Heelan (2004) within his phenomenological interpretation of the measurement problem in quantum mechanics claims:

"Husserl's *noetic-noematic union* of subject and object is an *entanglement* between the intentional subject and the emerging object – similar, perhaps, at this stage to Heidegger's *Dasein*."

The understanding of the nature of the "process" of measurement this allows a therapeutic "dissolution" or a metaphysical solution of the measurement problem.

**Conclusion**

Heidegger's phenomenology, as well as the second philosophy of Wittgenstein, can be understood as a normative pragmatism and even as a specific – normative – naturalism. It makes explicit the metaphysical presuppositions of the second philosophy of Wittgenstein. Both philosophies are appropriate for understanding quantum mechanics as a science of a new type, rejecting the metaphysical notion of reality, and for solving (in the case of Heidegger) or "dissolving" (in the case of Wittgenstein) the measurement problem in quantum mechanics.

Quantum concepts function rather as rules for forming a new reality, not as notions for describing a pre-existing metaphysical reality which is independent from the observer in the absolute sense.

The measurement problem in quantum mechanics has the same structure as the hard problem in the philosophy of mind, and can be reduced to the Wittgensteinian problem of rule-following. The "gap" between the potential possibilities theoretically described within quantum mechanics and the actualization of one of these possibilities is being closed pragmatically within a language game of a correct application of quantum mechanics playing the role of a Wittgensteinian rule. In the theoretical/metaphysical language of Heidegger, the language game is Dasein (or a realization of one of the potential possibilities contained in Dasein). The addition of this philosophical notion to the notional apparatus of quantum mechanics allows a theoretical solution to the measurement problem.